\documentstyle[aps,multicol]{revtex}

\begin{document}

\def\bb{\begin{equation}}
\def\ee{\end{equation}}
\def\cpp{c_{\perp}}
\def\cpr{c_{\parallel}}
\def\Gpr{G_{\parallel}}

\title{Photoemission in the system of linear chains : \\
application to PrBa$_2$Cu$_3$O$_7$ and 
La$_{\rm 2-x-y}$Nd$_{\rm y}$Sr$_{\rm x}$CuO$_4$}
\author{J.D. Lee, T. Mizokawa, and A. Fujimori}
\address{Department of Complexity Science and Engineering
and Department of physics, University of Tokyo, Bunkyo-ku, \\
Tokyo 113-0033, Japan}
\date{\today}
\maketitle

\begin{abstract}
Photoemission in the system of linear charged chains has been studied 
with the dipole matrix incorporated, from which its dependencies on 
the photoelectron momentum (${\bf k}$), photon polarization 
($\widehat{\bf E}$), and photon energy ($\omega$) 
can be explored. The used model is the three-dimensional 
array of noninteracting chains, which is so simple as to allow an
analytic approach. Motivations of the study are
for the doped CuO$_3$ chain in PrBa$_2$Cu$_3$O$_7$ and the doped
static stripe phase in La$_{\rm 2-x-y}$Nd$_{\rm y}$Sr$_{\rm x}$CuO$_4$.
The one-dimensional dispersion exhibiting  
spin-charge separation and its dependence on
the momentum perpendicular to the chain and its $\widehat{\bf E}$-dependency
as well are discussed in PrBa$_2$Cu$_3$O$_7$. 
For La$_{\rm 2-x-y}$Nd$_{\rm y}$Sr$_{\rm x}$CuO$_4$, 
the anomalous spectral distribution formed by
two sets of stripes perpendicular to each other 
is investigated.
The geometric effects led to by the dipole transition matrix 
including the interference of photocurrents from different chains are found
to change the simple one-dimensional feature drastically. 
We find these changes are consistent with experiment
for the chain system PrBa$_2$Cu$_3$O$_7$, but less satisfactory 
for the stripe phase in La$_{\rm 2-x-y}$Nd$_{\rm y}$Sr$_{\rm x}$CuO$_4$.
This means that in the stripe phase much two-dimensional
characters still exist unlike the chain system.

\end{abstract}


\begin{multicols}{2}

\section{Introduction}

Angle-resolved photoemission spectroscopy (ARPES) has become the most 
significant experimental tool for a search of the electronic
structure, especially in the high temperature superconducting
cuprates\cite{Shen95}. 
In many theoretical and experimental interpretations, 
the ARPES data is often compared to the one-electron spectral function 
$A({\bf k},\omega)$ or vice versa. For a more satisfactory 
description of the photoemission, i.e., if one is interested in 
various parameter dependencies or the line shapes of quasi-particle peaks,
one must go beyond the spectral function\cite{Hedin83}. For it,
we may need to consider the photoexcitation process governed 
by the dipole transition matrix\cite{Matsu97,Bansil} 
and the extrinsic losses of the photoelectron 
on its way out to and through the surface\cite{Hedin99,Lee99}.
In this paper, we give the study of transition matrix effects 
in the photoemission, while the extrinsic losses from the dynamical scattering
are not our concern because their contributions are in many cases smaller
by orders of magnitude.

The target systems are composed of charged linear chains, i.e., one-dimensional 
metals. One-dimensional metal has been known for the extreme
realization of electron correlation effects in that there are 
an exotic spin-charge separation and only
collective modes, not the single-particle excitation\cite{Tomo}.
It has opened another paradigm different from Fermi liquid
and is called Tomonaga-Luttinger liquid. In an actual situation, 
a one-dimensional metallic state is very unstable toward an insulating 
state or very marginal between metallic and insulating states
through charge and spin ordering.
Anyhow these one-dimensional chain systems have attracted considerable
attentions especially since the discovery of high $T_C$ superconductors.
Among the high $T_C$ cuprates, YBa$_2$Cu$_3$O$_7$ (YBCO) and its family
compounds have CuO$_3$ chain structure as well as CuO$_2$ plane.  
PrBa$_2$Cu$_3$O$_7$ (PBCO), among the YBCO family cuprates, does not support
superconductivity unlike other cuprates, which is suggested due to the
hole depletion in CuO$_2$ plane\cite{Take,Fehren}. Additional holes are 
thought to be doped in the CuO$_3$ chain, 
not in the CuO$_2$ plane. In such a sense,
PBCO can probably be an archetype of Tomonaga-Luttinger-type 
one-dimensional system,
which has revived much interests in the study of metallic chain system
within the study of high $T_C$ cuprates (PBCO actually has a band gap
opening possibly due to charge ordering in the chain\cite{Mizokawa99}).

On the other hand, another feasible usefulness of the metallic chain system
in high $T_C$ superconductor physics is in the so called stripe phase observed 
in Nd-substituted La$_{\rm 2-x-y}$Nd$_{\rm y}$Sr$_{\rm x}$CuO$_4$
(Nd-LSCO)\cite{Zaanen,Tranquada}. The stripe phase is due to the
formation of an ordered array of charged stripes which are also 
antiphase domain walls between antiferromagnetic ordered spins in the CuO$_2$ 
planes, which gives the periodicity-doubling for spins 
compared to charges.

Recent ARPES experiments for PBCO\cite{Mizokawa99} 
have shown actual evidence for one dimensionality
immersed in, but distinguishable from contributions from the two-dimensional 
CuO$_2$ planes. In experiments, two kinds of dispersive band features
are observed; one is from the undoped CuO$_2$ plane and the other
is the doped CuO$_3$ chain. Especially in the 
one-dimensional-like features, the separation between charge and 
spin excitation is also suggested. It is a stimulating result 
of Tomonaga-Luttinger liquid that the low energy physics is dominated by
uncoupled collective modes of charge and spin excitations,
also called holons and spinons.
In experiments, it is also found that,
in addition to the spin-charge separation, one-dimensional characters
(particularly spinon parts) are strongly dependent on the photon polarization
and at the same time on the momentum perpendicular to the chain.
The results for PBCO should be also relevant to other 
one-dimensional insulator, SrCuO$_2$, which has        
a weakly coupled double Cu-O chain. It is very intriguing
to find that the similar polarization and 
perpendicular momentum dependencies in ARPES are observed in SrCuO$_2$ 
as well\cite{Kim}. 
These observations could be hardly explained by 
a single Tomonaga-Luttinger chain only,
the additional effect probably needs incorporating, which motivated us  
to explore the dipole transition matrix related to the 
geometries of a chain system. 

The static stripe phases in Nd-LSCO
also lead to an interesting puzzle on the electronic structures
through ARPES observations\cite{Zhou}.
Prior to that, the neutron scattering experiments compellingly proposed that 
there should be the static stripe formation in Nd-LSCO
and besides, the charged stripes be  aligned with rotated
by $\pi/2$ between two adjacent planes\cite{Tranquada}.
The low binding energy electronic structure actually looks more or less 
one-dimensional in the sense that the spectral weight distribution 
is roughly understood as a superposition of two perpendicular chains.
However, the suppression of $d$-wave nodal signal along the Brillouin zone 
diagonal direction, particularly
around the $\Gamma$ point, could not be satisfactorily 
explained without additional physics. We suggest one necessary ingredient
could be from the dipole transition matrix including the interferences
from different stripes. Another noticeable point is that there have been 
no decisive evidences for Tomonaga-Luttinger liquid in the stripes phase
unlike chains in PBCO. Therefore, in our study for Nd-LSCO, 
although we consider the matrix effects from one-dimensional system, 
we assume the simple Fermi liquid for a single stripe having
well-defined Fermi point. Further, whether each stripe is a Fermi liquid
or a Tomonaga-Luttinger liquid does not affect the conclusion of 
the present work because the calculated polarization and photon energy 
dependencies of ARPES spectra come from the matrix element effects 
reflecting the interference of photoelectrons from different stripes.

In this work, we account for the dipole matrix effects for the
semi-infinite ($z>0$) three-dimensional array of metallic chains, 
whose Green's function is considered. The slight photoelectron damping
is taken into account in the simplest way.
It is assumed that 
all the excitations including both charge and spin occur
along the chain-direction, that is, the energy scale for the
perpendicular excitation is very high. This means that we neglect 
the interchain (interstripe) interactions 
such as hybridization or Coulomb interaction,
which may exist in the actual system and also 
the fluctuation of chain or stripe perpendicular to it,
which could be important in a dynamical stripe phase. 
We think of two configurations.
First, we simply model the three-dimensional array of chains
for PBCO. On the other hand, 
for Nd-LSCO, we slightly extend the model
and consider the three-dimensional chain array perpendicularly
crossed in adjacent planes. For both configurations,
the model is analytically tractable.
The resulting spectra is determined 
by complicated interferences of chain contributions and thus 
the dipole matrix effects are geometric. The effects naturally give
${\bf k}$- and $\widehat{\bf E}$-dependent photoemission spectra,
which is found consistent with the recent experiments.

The outline of this paper is as follows. 
First we discuss the basic theory and the suitable starting point
to account for the dipole transition matrix in Sec. II.
In Sec. III, we investigate the model for PBCO and discuss its results.
We can see how the spectral function of Luttinger liquid changes
by the transition matrix. Such changes are also found
quite consistent with experiments. In Sec. IV, we extend and
constitute the model for Nd-LSCO and obtain the
spectral weight distribution through the similar type of calculation.
The calculation is also more or less consistent with 
the low energy weight distribution
in experiments (here we assume {\it a priori} metallic stripes)
in a limited case, which is from the interference
among chains. In the final Sec. V, we present the concluding 
remarks and give the outlook.

\section{Basic Formalism}

To account for the dipole transition matrix in the
photoemission spectral intensity, we follow Pendrey's formula\cite{Pendrey}.
Its benefit is to treat only one-particle Green's function without solving for
any static stationary states. From Pendrey's the photoemission
intensity $I({\bf k},E,\omega)$ is
$$
I({\bf k},E,\omega)=-\frac{1}{\pi}{\rm Im}M({\bf k},E,\omega),
$$
\bb\label{II1}
M({\bf k},E,\omega)\equiv
\langle {\bf k}|G^+(E+\omega)\Delta G^+(E)\Delta^{\dagger}G^-(E+\omega)
|{\bf k}\rangle,
\ee
where $G^{\pm}$ is the retarded or advanced one-particle Green's function
and $|{\bf k}\rangle$ the photoelectron state. The photoelectron state could
be represented as a plane wave in the lowest approximation, even if 
the real photoelectron wave function may be damped through the scattering
with possible excitations in the solid, i.e., could be quite different from
a plane wave except for in the very high energy region. In our study,
assuming the solid occupies $z>0$ space, we take into account the damping 
in the simplest way by 
taking the photoelectron wave as a plane wave in the in-plane 
direction and a damped wave in the $z$-direction. This would not be 
a very good representation due to the anisotropy inherent in the system
even in the in-plane direction\cite{Hedin98}.
We would expect and show, nevertheless, this could catch
the essential physics as it should have.
Then in Eq.(\ref{II1}) we replace $|{\bf k}\rangle$ by
$|{\bf \tilde{k}}\rangle$, which is given as
\bb\label{II2}
\langle{\bf r}|{\bf \tilde{k}}\rangle
=\frac{1}{(2\pi)^{3/2}}
 e^{-i{\bf K}\cdot{\bf R}}[\theta(z)e^{-i\tilde{k}_z^{\ast}z}
                          +\theta(-z)e^{-ik_z z}].
\ee
$\tilde{k}_z$ is a complex number, $\tilde{k}_z=\sqrt{k_z^2+i\Gamma}$,
and thus the photoelectron gets damped. 
It is worth noting that a damped photoelectron state is the formal solution
of the one-electron Schr\"{o}donger equation
\bb\label{II3}
(\epsilon_{\bf k}-{\cal H}_0-\Sigma)|{\bf k}\rangle=0,
\ee
where $\Sigma$ is the self-energy operator involving
the photoelectron-solid interaction. Equation (\ref{II2}) can be understood
from the first approximation for its self-energy, 
$\Sigma=-i\Gamma\theta(z)$. 
If we insert the plane wave states in Eq.(\ref{II1}), the photoemission
matrix $M({\bf k},E,\omega)$ is dissolved as
\begin{eqnarray}\label{II4}
M({\bf k},E,\omega)&=&\int d{\bf k}^{\prime}d{\bf k}^{\prime\prime}
d{\bf k}^{\prime\prime\prime}d{\bf k}^{\prime\prime\prime\prime}
\langle{\bf \tilde{k}}|G_2^+|{\bf k}^{\prime}\rangle\langle{\bf k}^{\prime}|
\Delta|{\bf k}^{\prime\prime}\rangle
\\ \nonumber
& &\times \langle{\bf k}^{\prime\prime}|
G_1^+|{\bf k}^{\prime\prime\prime}\rangle\langle{\bf k}^{\prime\prime\prime}|
\Delta^{\dagger}|{\bf k}^{\prime\prime\prime\prime}\rangle
\langle{\bf k}^{\prime\prime\prime\prime}|G_2^-|{\bf \tilde{k}}\rangle,
\end{eqnarray}
where $G_1$ and $G_2$ denote the one-particle Green's functions 
at $E$ and $E+\omega$, respectively, and 
$\Delta=\widehat{\bf E}\cdot{\bf p}$, we would get,
assuming the linear polarized photon,
\bb\label{II5}
\langle{\bf k}^{\prime}|\Delta|{\bf k}^{\prime\prime}\rangle
=(\widehat{\bf E}\cdot{\bf k}^{\prime})
 \delta({\bf k}^{\prime}-{\bf k}^{\prime\prime}), \ \
{\bf k}=({\bf K},k_z).
\ee

\section{Photoemission in the system of chains : 
         for P\lowercase{r}B\lowercase{a}$_2$C\lowercase{u}$_3$O$_7$}

The low energy features in ARPES for PBCO are expected to be governed by the
CuO$_3$ chain. In a formula unit, $R$Ba$_2$Cu$_3$O$_7$ 
($R$=rare earth) has pairs of CuO$_2$ planes between CuO$_3$ chain layers.
Among those compounds, for the case of PrBa$_2$Cu$_3$O$_7$, 
various experimental\cite{Take} and theoretical studies\cite{Fehren}
suggests that the CuO$_2$ plane should not be doped, 
but the CuO$_3$ chain doped.
In this study, we do not consider the undoped CuO$_2$ plane
since its contributions will be pushed to high binding energies.

The chains are along the $y$-direction ($b$-axis). Then the formal 
Green's function of the system in the real space representation
can be written as
\bb\label{III1}
G^{\pm}({\bf r},{\bf r}^{\prime})=\sum_s\sum_{m{m^{\prime}}}
         \frac{\psi_s^{mm^{\prime}}({\bf r})
               {\psi_s^{mm^{\prime}}}^{\ast}({\bf r}^{\prime})}
              {E-\epsilon_s\pm i0^+},
\ee
where $\psi_s^{mm^{\prime}}({\bf r})$ are eigenstates of the system,
$\psi_s^{mm^{\prime}}({\bf r})=\psi_s(y)\chi(x-\cpr m)\chi(z-\cpp m^{\prime})$.
Here $\psi_s(y)$ is assumed to include both charge and spin degrees 
of freedom, while in the direction perpendicular to the chain
the minimum excitation energy is assumed to be so high that
$\chi(x)$ or $\chi(z)$ is taken the lowest localized one-dimensional state.
Besides, we assume the overlap between the chains is very small.
Then, if we rewrite Eq.(\ref{III1}) in a more explicit way, 
\begin{eqnarray}\label{III2}
G^{\pm}({\bf r},{\bf r}^{\prime})&=&\bar{G}^{\pm}(y-y^{\prime})
              \sum_{m=-\infty}^{\infty}\chi(x-\cpr m)
              \chi^{\ast}(x^{\prime}-\cpr m)
\nonumber \\
&\times&
              \sum_{m^{\prime}=0}^{\infty}
                  \chi(z-\cpp m^{\prime}-\frac{\cpp}{2})
                  \chi^{\ast}(z^{\prime}-\cpp m^{\prime}-\frac{\cpp}{2}),
\nonumber \\
\end{eqnarray}
where we have assumed that $z=0$ is the surface 
and the first chain layer is positioned at $z=\frac{\cpp}{2}$.
$\bar{G}^{\pm}(y-y^{\prime})$ is an ordinary single-particle
Green's function in the Luttinger liquid including both
charge and spin degrees of freedom. In the Luttinger model,
the single-particle Green's function and its property
are already well known\cite{Meden}. This trivial factorization of
the Luttinger Green's function is thanks to the complete suppression
of the perpendicular excitation by the starting assumption.
It should be also noted that, for the interchain distances,   
experiments give $c_a\sim c_b=\cpr$ and 
$\cpp\sim 3\cpr$\cite{Grevin}.

Now we see $\langle{\bf \tilde{k}}|G|{\bf k}^{\prime}\rangle$ is
\begin{eqnarray}\label{III3}
\langle{\bf \tilde{k}}|G|{\bf k}^{\prime}\rangle
&=&\bar{G}(k_y)\delta(k_y-k_y^{\prime})
 \sum_{m=0}^{\infty}\tilde{\chi}^m(k_z)[\bar{\chi}^m(k_z^{\prime})]^{\ast}
\nonumber \\
&\times&
 \chi(k_x)\chi^{\ast}(k_x^{\prime})
 \left[\frac{1}{\cpr}\sum_{\Gpr}\delta(k_x-k_x^{\prime}+\Gpr)\right],
\end{eqnarray}
where $\Gpr=2n\pi/\cpr$ is the in-plane reciprocal lattice vector 
due to the periodic lattice. For the explicit calculation, we propose
a suitable exponential form for $\chi(x)$ (or $\chi(z)$),
\bb\label{III4} 
\chi(x)=\frac{1}{\sqrt{a}}e^{-|x|/a}, \ \ \
\chi(k_x)=\sqrt{\frac{2a}{\pi}}\frac{1}{1+k_x^2a^2},
\ee
and additionally, 
\begin{eqnarray*}
\tilde{\chi}^m(k_z) 
               &=&\frac{1}{\sqrt{2\pi a}}\frac{2/a}{1/a^2+\tilde{k}_z^2}
                  e^{i\tilde{k}_z(\cpp m+\cpp/2)}
\nonumber \\
               &-&\frac{1}{\sqrt{2\pi a}}
                 \left[\frac{1}{1/a+i\tilde{k}_z}-\frac{1}{1/a+ik_z}\right]
                  e^{-\frac{1}{a}(\cpp m+\cpp/2)},
\end{eqnarray*}
\bb\label{III5}
\bar{\chi}^m(k_z)=\chi(k_z)e^{ik_z(\cpp m+\cpp/2)}.
\ee
Here we basically assume $\cpr,\cpp \gg a$.
In Eq.(\ref{III5}), $\tilde{\chi}^m(k_z)$ is reduced into 
$\bar{\chi}^m(k_z)$ when replacing $\tilde{k}_z$ by $k_z$.
Using Eqs.(\ref{III3}),(\ref{III4}), and (\ref{III5}), 
we get the photoemission matrix $M({\bf k},E,\omega)$, 
aside from irrelevant constants,
\begin{eqnarray}\label{III6}
M({\bf k},E,\omega)&=&|\bar{G}_2(k_y)|^2\bar{G}_1^+(k_y)|\chi(k_x)|^2
\\ \nonumber
                   &\times&
                   \sum_m\sum_{m^{\prime}}\sum_{m^{\prime\prime}}
                   \tilde{\chi}^m(k_z)
                   [{\tilde{\chi}^{m^{\prime\prime}}}(k_z)]^{\ast}
\\ \nonumber
                   &\times&
                   \int dk_z^{\prime}dk_z^{\prime\prime}
                   \sum_{\Gpr}\sum_{\Gpr^{\prime}}
                   |\chi(k_x+\Gpr)|^2|\chi(k_x+\Gpr^{\prime})|^2
\\ \nonumber
                   & & \ \ \ \ \ \ \ \ \times
                   [\varepsilon_x(k_x+\Gpr)+\varepsilon_yk_y
                    +\varepsilon_zk_z^{\prime}]
\\ \nonumber
                   & & \ \ \ \ \ \ \ \ \times                          
                   [\varepsilon_x(k_x+\Gpr^{\prime})+\varepsilon_yk_y
                    +\varepsilon_zk_z^{\prime\prime}]
\\ \nonumber
                   & & \ \ \ \ \ \ \ \ \times 
                   [\bar{\chi}^{m}(k_z^{\prime})]^{\ast}
                   \bar{\chi}^{m^{\prime}}(k_z^{\prime})
                   [\bar{\chi}^{m^{\prime}}(k_z^{\prime\prime})]^{\ast}
                   \bar{\chi}^{m^{\prime\prime}}(k_z^{\prime\prime}),
\end{eqnarray}
where $\widehat{\bf E}=(\varepsilon_x,\varepsilon_y,\varepsilon_z)$ and 
$m$, $m^{\prime}$, and $m^{\prime\prime}$ run from 0 to $+\infty$.
Equation (\ref{III6}) involves two types of summations with respect to
$\Gpr$. The summation for the reciprocal lattice vector can be done 
by use of 
\bb\label{III7}
\sum_{\Gpr}e^{i\Gpr x}=\cpr\sum_{m=-\infty}^{\infty}\delta(x-\cpr m).
\ee
First, using Eq.(\ref{III7}), let us note and define 
\begin{eqnarray}\label{III8}
\sum_{\Gpr}|\chi(k_x+\Gpr)|^2&=&\frac{2a}{\pi}\sum_{\Gpr}
                            \frac{1}{[1+(k_x+\Gpr)^2a^2]^2}
\nonumber \\
   &=&\frac{\cpr}{2\pi}\sum_m e^{ik_x\cpr m}
                    [\cpr|m|/a+1]e^{-\cpr|m|/a}
\nonumber \\
   &\equiv&\xi(k_x),
\end{eqnarray}
and in the same way,
\begin{eqnarray}\label{III9}
\sum_{\Gpr}(k_x+\Gpr)|\chi(k_x+\Gpr)|^2&=&\frac{2a}{\pi}\sum_{\Gpr}
                    \frac{k_x+\Gpr}{[1+(k_x+\Gpr)^2a^2]^2}
\nonumber \\
 &=&-\frac{i}{2\pi}\frac{\cpr^2}{a^2}
      \sum_{m\neq 0}e^{ik_x \cpr m}me^{-\cpr|m|/a}
\nonumber \\
&\equiv&\zeta(k_x).
\end{eqnarray}
For the $x$-direction, the lattice is so infinitely periodic that 
$m$ runs from $-\infty$ to $+\infty$, which should be distinguished
from the $z$-direction. In an actual calculation of
$\xi(k_x)$ and $\zeta(k_x)$, it will be enough to have the first
few terms of the infinite series because we have a well-defined 
small expansion parameter $e^{-\cpr/a}$.
For a further calculation of Eq.(\ref{III6}), 
we feel like getting a simpler form of $\tilde{\chi}^m(k_z)$
than in Eq.(\ref{III5}). If we think of a small damping case
($k_z\gg\Gamma$), we can simplify Eq.(\ref{III5}) into
\bb\label{III10}
\tilde{\chi}^m(k_z)\approx
\chi(k_z)e^{-{\rm Im}\tilde{k}_z(\cpp m+\cpp/2)}e^{ik_z(\cpp m+\cpp/2)},
\ee
and ${\rm Im}\tilde{k}_z\approx\Gamma/2|k_z|$. On the other hand, 
we can readily perform the remaining integrals for $k_x^{\prime}$ and
$k_x^{\prime\prime}$,
\begin{eqnarray}\label{III11}
\int dk_z[\bar{\chi}^m(k_z)]^{\ast}\bar{\chi}^{m^{\prime}}(k_z)
&=&\frac{2}{\pi a^3}
   \int dk_z\frac{e^{-ik_z\cpp(m-m^{\prime})}}{(1/a^2+k_z^2)^2}
\nonumber \\
&=&[\cpp|m-m^{\prime}|/a+1]e^{-\cpp|m-m^{\prime}|/a}
\nonumber\\
&\equiv&\bar{\xi}(m-m^{\prime}),
\end{eqnarray}
and also in the same way,
\begin{eqnarray}\label{III12}
\int dk_z k_z[\bar{\chi}^m(k_z)]^{\ast}\bar{\chi}^{m^{\prime}}(k_z)
&=&\frac{2}{\pi a^3}
   \int dk_z k_z\frac{e^{-ik_z\cpp(m-m^{\prime})}}{(1/a^2+k_z^2)^2}
\nonumber \\
&=&-\frac{i}{a}\frac{\cpp}{a}(m-m^{\prime})e^{-\cpp|m-m^{\prime}|/a},
\nonumber \\
&\equiv&\bar{\zeta}(m-m^{\prime}).
\end{eqnarray}
From Eqs.(\ref{III8})-(\ref{III12}), we can reexpress $M({\bf k},E,\omega)$
into a form including only the summations for $m$'s,
\begin{eqnarray}\label{III13}
M({\bf k},E,\omega)&=&|\bar{G}_2(k_y)|^2\bar{G}_1^+(k_y)
            |\chi(k_x)|^2|\chi(k_z)|^2e^{-{\rm Im}\tilde{k}_z \cpp}
\nonumber \\
           &\times& \sum_m\sum_{m^{\prime}}\sum_{m^{\prime\prime}}
            e^{-{\rm Im}\tilde{k}_z \cpp(m+m^{\prime\prime})}
            e^{ik_z\cpp(m-m^{\prime\prime})}
\nonumber \\
            &\times&
            [\varepsilon_x^2\zeta(k_x)^2\bar{\xi}(m-m^{\prime})
             \bar{\xi}(m^{\prime}-m^{\prime\prime})
\nonumber \\   
            & &
            +\varepsilon_y^2k_y^2\xi(k_x)^2\bar{\xi}(m-m^{\prime})
             \bar{\xi}(m^{\prime}-m^{\prime\prime})
\nonumber \\
            & & 
            +\varepsilon_z^2\xi(k_x)^2\bar{\zeta}(m-m^{\prime})
             \bar{\zeta}(m^{\prime}-m^{\prime\prime})
\nonumber \\
            & &
            +2\varepsilon_x\varepsilon_y k_y\xi(k_x)\zeta(k_x)
             \bar{\xi}(m-m^{\prime})\bar{\xi}(m^{\prime}-m^{\prime\prime})
\nonumber \\
            & & 
            +\varepsilon_y\varepsilon_zk_y\xi(k_x)^2\bar{\xi}(m-m^{\prime})
             \bar{\zeta}(m^{\prime}-m^{\prime\prime})
\nonumber \\
            & &
            +\varepsilon_y\varepsilon_zk_y\xi(k_x)^2
             \bar{\zeta}(m-m^{\prime})
             \bar{\xi}(m^{\prime}-m^{\prime\prime})
\nonumber \\
            & & 
            +\varepsilon_z\varepsilon_x\xi(k_x)\zeta(k_x)
             \bar{\xi}(m-m^{\prime})\bar{\zeta}(m^{\prime}-m^{\prime\prime})
\nonumber \\
            & &
            +\varepsilon_z\varepsilon_x\xi(k_x)\zeta(k_x)
             \bar{\zeta}(m-m^{\prime})
             \bar{\xi}(m^{\prime}-m^{\prime\prime})].
\nonumber \\
\end{eqnarray}
This looks still formidable complicated summations, but 
we can do a systematic summation thanks to a small parameter 
$e^{-\cpp/a}$ (or $e^{-\cpr/a}$, note $\cpp\sim3\cpr$).
In order to retain up to the lowest order of $e^{-\cpp/a}$,
we can do a summation keeping $|m-m^{\prime}|\leq 1$ and 
$|m^{\prime}-m^{\prime\prime}|\leq 1$. This is in the end same order
of approximation in $\xi(k_x)$ and $\zeta(k_x)$ up to $e^{-3\cpr/a}$.
That is, we have from Eqs.(\ref{III8}) and (\ref{III9}),
\begin{eqnarray}\label{III14}
\xi(k_x)&\approx&\frac{\cpr}{2\pi}[1+2\cos(k_x\cpr)(\cpr/a+1)e^{-\cpr/a}
\nonumber \\
           & & \ \ \ \ \ \ \ +2\cos(2k_x\cpr)(2\cpr/a+1)e^{-2\cpr/a}
\nonumber \\
           & & \ \ \ \ \ \ \ +2\cos(3k_x\cpr)(3\cpr/a+1)e^{-3\cpr/a}],
\end{eqnarray}
\begin{eqnarray}\label{III15}
\zeta(k_x)&\approx&\frac{1}{\pi}\frac{\cpr^2}{a^2}
             [\sin(k_x\cpr)e^{-\cpr/a}+2\sin(2k_x\cpr)e^{-2\cpr/a}
\nonumber \\
           & & \ \ \ \ \ \ +3\sin(3k_x\cpr)e^{-3\cpr/a}].
\end{eqnarray}
We then have three kinds of summations in Eq.(\ref{III13}); we define those
as $\alpha(k_z)$, $\beta(k_z)$, and $\gamma(k_z)$.
\begin{eqnarray}\label{III16}
\alpha(k_z)&\times&(1-e^{-2{\rm Im}\tilde{k}_z\cpp})
\nonumber \\
&\equiv&\sum_m\sum_{m^{\prime}}\sum_{m^{\prime\prime}}
e^{-{\rm Im}\tilde{k}_z \cpp(m+m^{\prime\prime})}
e^{ik_z \cpp(m-m^{\prime\prime})}
\nonumber \\
& & \ \ \ \ \times
\bar{\xi}(m-m^{\prime})\bar{\xi}(m^{\prime}-m^{\prime\prime})
(1-e^{-2{\rm Im}\tilde{k}_z\cpp})
\nonumber \\
&\approx&1+(\cpp/a+1)^2e^{-2\cpp/a}(1+e^{-2{\rm Im}\tilde{k}_z \cpp})
\nonumber \\
   & &
   +4\cos(k_z \cpp)(\cpp/a+1)e^{-\cpp/a}e^{-{\rm Im}\tilde{k}_z \cpp}
\nonumber \\
   & &
   +2\cos(2k_z \cpp)(\cpp/a+1)^2e^{-2\cpp/a}e^{-2{\rm Im}\tilde{k}_z \cpp},
\end{eqnarray}
\begin{eqnarray}\label{III17}
\beta(k_z)&\times&(1-e^{-2{\rm Im}\tilde{k}_z\cpp})
\nonumber \\
&\equiv&\sum_m\sum_{m^{\prime}}\sum_{m^{\prime\prime}}
e^{-{\rm Im}\tilde{k}_z \cpp(m+m^{\prime\prime})}
e^{ik_z \cpp(m-m^{\prime\prime})}
\nonumber \\
& & \ \ \ \ \times
\bar{\zeta}(m-m^{\prime})\bar{\zeta}(m^{\prime}-m^{\prime\prime})
(1-e^{-2{\rm Im}\tilde{k}_z\cpp})
\nonumber \\
&\approx&\frac{1}{a^2}(\cpp/a)^2e^{-2\cpp/a}(1+e^{-2{\rm Im}\tilde{k}_z \cpp})
\nonumber \\
&-&\frac{2}{a^2}\cos(2k_z \cpp)(\cpp/a)^2e^{-2\cpp/a}
e^{-2{\rm Im}\tilde{k}_z \cpp},
\end{eqnarray}
\begin{eqnarray}\label{III18}
\gamma(k_z)&\times&(1-e^{-2{\rm Im}\tilde{k}_z\cpp})
\nonumber \\
&\equiv&{\rm Re}\sum_m\sum_{m^{\prime}}\sum_{m^{\prime\prime}}
e^{-{\rm Im}\tilde{k}_z \cpp(m+m^{\prime\prime})}
e^{ik_z \cpp(m-m^{\prime\prime})}
\nonumber \\
& & \ \ \ \ \ \ \ \ \times     
\bar{\xi}(m-m^{\prime})\bar{\zeta}(m^{\prime}-m^{\prime\prime})
(1-e^{-2{\rm Im}\tilde{k}_z\cpp})
\nonumber \\
&\approx&\frac{2}{a}\sin(2k_z \cpp)(\cpp/a+1)(\cpp/a)
   e^{-2\cpp/a}e^{-2{\rm Im}\tilde{k}_z\cpp}
\nonumber \\
&+&\frac{2}{a}\sin(k_z \cpp)(\cpp/a)e^{-\cpp/a}e^{-{\rm Im}\tilde{k}_z\cpp}.
\end{eqnarray}
Now, all the summation parts of Eq.(\ref{III13}) are found to be real
as it should be. Therefore, the photoemission intensity 
$I({\bf k},E,\omega)$ will be proportional to 
$\left[-\frac{1}{\pi}{\rm Im}\bar{G}_1^+(k_y)\right]$. Another
factor related to the chain degree of freedom is 
$|\bar{G}_2(k_y)|^2(=[{\rm Re}\bar{G}_2(k_y)]^2+[{\rm Im}\bar{G}_2(k_y)]^2)$, 
which gives the structure in the final states. Since $E+\omega\gg E$,
we can assume the final states form a structureless smooth continuum.
We do not have to make the problem difficult unnecessarily.

Then we can find an analytic solution of the problem. The photoemission
intensity $I({\bf k},E,\omega)$ is 
\begin{eqnarray}\label{III19}
I({\bf k},E,\omega)&\propto&
                    \left[-\frac{1}{\pi}{\rm Im}\bar{G}_1^+(k_y)\right]
                    |\chi(k_x)|^2|\chi(k_z)|^2
                    e^{-{\rm Im}\tilde{k}_z \cpp}
\nonumber \\
            &\times&
            [\varepsilon_x^2\zeta(k_x)^2\alpha(k_z)
            +\varepsilon_y^2k_y^2\xi(k_x)^2\alpha(k_z)
\nonumber \\   
           & & 
            +\varepsilon_z^2\xi(k_x)^2\beta(k_z)
            +2\varepsilon_x\varepsilon_y k_y\xi(k_x)\zeta(k_x)\alpha(k_z)
\nonumber \\
           & &
            +2\varepsilon_y\varepsilon_zk_y\xi(k_x)^2\gamma(k_z)
            +2\varepsilon_z\varepsilon_x\xi(k_x)\zeta(k_x)\gamma(k_z)],
\nonumber \\
\end{eqnarray} 
where we note $k_z=\sqrt{2(E+\omega)-k_x^2-k_y^2}$.

\begin{figure}
\vspace*{10.5cm}
\includegraphics{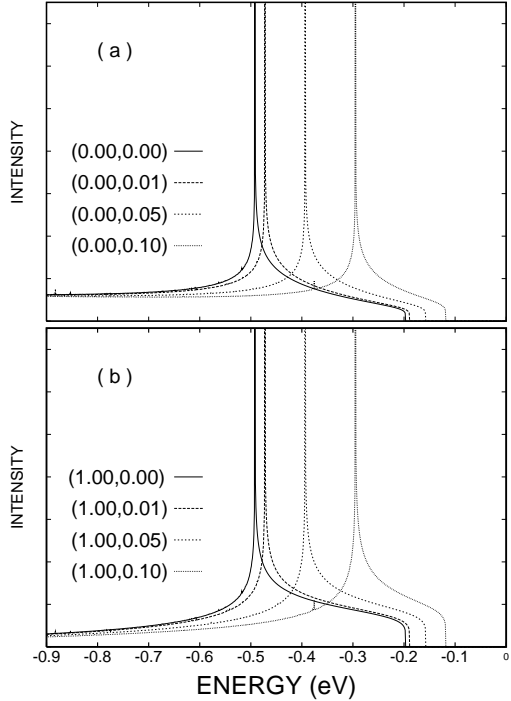} 
\caption{Photoemission spectra in the chain direction ($\parallel k_y$)
for $\widehat{\bf E}\parallel xz$.
It is seen that the strong $k_x$-dependencies are inherent in the spectra.
In (a), the spectra along $(0,0)\to(0,k_y)$ and in (b),
the spectra along $(1,0)\to(1,k_y)$ are provided. 
}
\label{fig:Exz}
\end{figure}

For an actual calculation and its comparison with experiment, 
the reference experiment should be Mizokawa {\it et al.}'s\cite{Mizokawa99}.
In the experiment, the Fermi energy (the width of occupied band)
is estimated about 0.2 eV and the Fermi wave vector $k_f=0.25\frac{\pi}{\cpr}$.
In theory, the Fermi energy $\epsilon_f$
is given by the linear dispersion within Luttinger model such that 
$\epsilon_f=v_fk_f$. In the spectral function of Luttinger model, 
i.e., in $\left[-\frac{1}{\pi}{\rm Im}\bar{G}_1^+(k_y)\right]$, 
we see two features corresponding to the collective charge (holon) 
and spin (spinon) excitations.

\begin{figure}
\vspace*{10.5cm}
\includegraphics{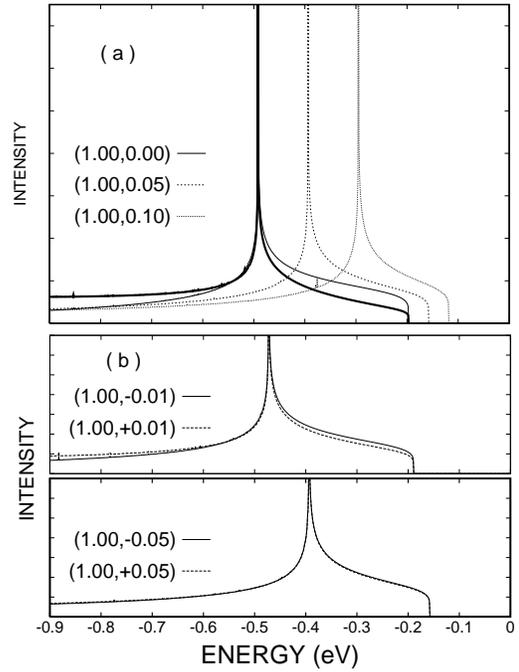}
\caption{Photoemission spectra for $\widehat{\bf E}\parallel yz$,
where ${\bf E}$ has a component parallel to the chain.
In (a), the spectra along $(1,0)\to(1,k_y)$ and in (b),
the spectra along $(1,-k_y)\to(1,k_y)$ are provided. The spectra
at $(0,0)$ are also given in (a) as a thick solid line.
}
\label{fig:Eyz}
\end{figure}

\noindent
The character of features generally
rely on the interaction property through the singularity index 
$\alpha$. If $\alpha<1/2$, there happens a diverging sharp peak at 
the spinon onset, but if $\alpha>1/2$, a converging threshold.
$\alpha\sim 0.65$ is estimated from optical study 
of PrBa$_2$Cu$_4$O$_8$\cite{Take00}. Here we employ
this value for PBCO.
The charge contribution is peaked at $-v_f^c(k_f-k_y)$ 
and the spin excitation at
$-v_f(k_f-k_y)$, where we are having the spin-independent interaction in mind.
$v_f^c$ for the holon is taken a little larger
as $v_f^c/v_f\sim 2.5$ to be consistent with 
qualitative behaviors of experiment\cite{Den}.
The natural energy unit in theory is $(\pi/\cpr)^2$, whose value is
about 5.2 eV from $\cpr\sim 3.8$\AA.
From the dispersion, we put $\epsilon_f=v_fk_f$, 
$v_f\sim 0.15\frac{\pi}{\cpr}$ 
and $\epsilon_f\sim 0.038(\pi/\cpr)^2$. 
The used photon energy in the experiment
is 29 eV and $\sim 145\epsilon_f$, from which we take in the calculation
the photon energy $\omega\sim 145 \epsilon_f\sim 5.5 (\pi/\cpr)^2$.
However the calculation is found quite robust with the parameters.
The next thing to be considered is a constant damping parameter
$\Gamma$. We take $\Gamma=0.1(\pi/\cpr)^2$. This corresponds to 
the mean free path $\lambda\sim \frac{v}{2\Gamma}\sim 
\frac{\sqrt{2\omega}}{2\Gamma}\sim 5.3\cpr$. Thinking of $\cpp\sim 3\cpr$
and that the first layer is at $\frac{\cpp}{2}$, $\lambda$ amounts to
governing one or two layers. This looks more or less acceptable.
In the calculation, we take $\cpr/a=2.5$, i.e., $\cpp/a=7.5$.

\begin{figure}
\vspace*{5.cm}
\includegraphics{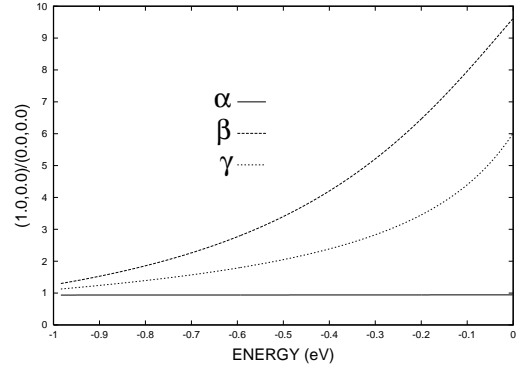}
\caption{The ratios of $\alpha(k_z)$, $\beta(k_z)$, and $\gamma(k_z)$
for two simple cases (1,0) and (0,0) are given, i.e.,  
$\alpha=\frac{\alpha(k_z,k_x=1,k_y=0)}{\alpha(k_z,k_x=0,k_y=0)}$
and same as for $\beta$ and $\gamma$.
}
\label{fig:beta}
\end{figure}

Due to the one-dimensionality itself, the chain system is expected 
to show a strong polarization dependency. In fact, this is a useful
clue in finding one-dimensional character submerged in
two-dimensional properties.
In Fig.\ref{fig:Exz}, the photoemission spectra for the 
photon polarization $\widehat{\bf E}\parallel xz$, more explicitly
$\widehat{\bf E}=(\frac{1}{\sqrt{2}},0,\frac{1}{\sqrt{2}})$,
are shown.
In the figure, we attribute the high binding energy structure,
the sharp peak dispersing between $\sim-0.5$ eV and $\sim-0.3$ eV
to the charge excitation (holon) and the low binding energy structure,
the threshold at $\sim-0.2$ eV
to the spin excitation (spinon). When $k_x=0$ (Fig.\ref{fig:Exz}(a)),
the spinon part is highly suppressed, while when $k_x=1$
(Fig.\ref{fig:Exz}(b)), the spinon part is much enhanced.
In the experiment\cite{Mizokawa99}, for $k_x=0$ or small $k_x$'s, 
the spinon branch has not been observed and for $k_x\sim 1$, the spinon
has been so enhanced that it can be observed. Therefore,
these theoretical and experimental observations has been found in
good agreement and consistent with each other. 
It need to be noticed that PBCO is an insulator due to a charge density
wave (CDW) gap opening at $k_y\sim 0.25\frac{\pi}{\cpr}$, 
while the Luttinger model signifies in principle a metallic system.
In the study, we provide the results for $k_y$'s away from $k_f$
because the spin-charge separation is seen only for small $|k_y|$
and disappears for $k_y\sim k_f$. 
Further agreement can be found for 
$\widehat{\bf E}\parallel yz$, actually $\widehat{\bf E}=(0,\frac{1}{\sqrt{2}},
\frac{1}{\sqrt{2}})$ in Fig.\ref{fig:Eyz}. 
In the experiment\cite{Mizokawa99}, the spectra have been suppressed for 
small $k_x$, but comparable to the configuration of 
$\widehat{\bf E}\parallel xz$ when $k_x\sim 1$. This is well understood
from our theory as shown in Fig.\ref{fig:Eyz}(a). 
Comparison of the simplest cases (1,0) and (0,0) for both polarizations
(Figs.\ref{fig:Exz} and \ref{fig:Eyz}) shows the 
$k_x$(perpendicular momentum)-dependency clearly,
whose behaviors are determined by the third term in Eq.(\ref{III19}),
$\xi(k_x)^2\beta(k_z)$. It is shown in Fig.\ref{fig:beta}
that $\beta(k_z)$ could reinforce or suppress the spinon parts 
depending on the momentum perpendicular to the chain, i.e., $k_x$.
In the configuration of $\widehat{\bf E}\parallel yz$, 
the ${\bf E}$ has a component parallel to the chain,
so the symmetry of the system, when $k_y\leftrightarrow -k_y$, breaks.
Indeed,  the phase space showing the stronger spinon contribution is
along $(1,0)\to(1,-k_y)$ rather than $(1,0)\to(1,k_y)$ (for $k_y>0$).
Figure \ref{fig:Eyz}(b) shows this asymmetry also successfully 
at least in a qualitative picture. In the calculations, the asymmetry
comes from $k_y\xi(k_x)^2\gamma(k_z)$ in Eq.(\ref{III19}).
The obtained asymmetry is in fact
weaker than the experiment and rapidly fades away as $|k_y|$ increases.
In the figures, the absolute intensity
variations with respect to several ${\bf k}$'s would not
be very reliable here because of many simplifications.
Nevertheless, our model of the 1/4-filled 
three-dimensional chain array is found to explain very well 
the one-dimensional behaviors, especially the relative charge 
and spin intensities, reflected in the recent ARPES data.

Another one-dimensional insulator SrCuO$_2$ also has Cu-O chains.
Even if there is a finite interchain coupling, it is an order of magnitude
weaker than the intrachain coupling, which makes SrCuO$_2$
a one-dimensional compound.
Aside from several details, we find that the present calculations 
should be relevant to the SrCuO$_2$ system as well. In fact, the
ARPES experiment on SrCuO$_2$ shows that the qualitative behaviors
are very similar to the PBCO case (See Figs. 4 and 5 in Ref.\cite{Kim}) 
in that the spinon bands are more apparent in the case of large ($\sim 1$)
momentum perpendicular to the chain.

\section{Photoemission in the system of chains :
         For L\lowercase{a}\lowercase{$_{\rm 2-x-y}$}
             N\lowercase{d}\lowercase{$_{\rm y}$}
             S\lowercase{r}\lowercase{$_{\rm x}$}C\lowercase{u}O$_4$}

\begin{figure}
\vspace*{3.5cm}
\includegraphics{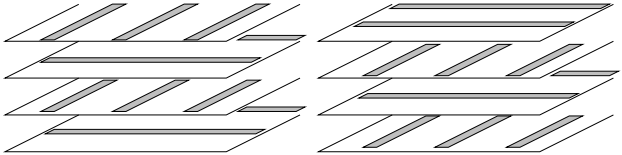}
\caption{Schematic view of the static stripe phase observed in Nd-LSCO.
In the study, the stripe would be modeled as a linear charged chain.
In an actual experimental situation, both configurations have been
equally mixed.
}
\label{fig:NdLSCO-stripe}
\end{figure}

In the so called stripe phase, there happens the one-dimensional
spin-charge modulations in the two-dimensional CuO$_2$ planes.
The stripe phase has attracted much attention in connection with 
its implication to high $T_C$ superconductivity since it was 
proposed from neutron scattering in Nd-substituted
La$_{\rm 2-x-y}$Nd$_{\rm y}$Sr$_{\rm x}$CuO$_4$ (Nd-LSCO)\cite{Tranquada}.
It has been also reported that the charge transport in Nd-LSCO is observed 
to be one-dimensional, also consistent with the static stripe phase 
formation\cite{Noda}. 
Similar stripe signatures are observed in La$_{\rm 2-x}$Sr$_{\rm x}$CuO$_4$ 
(LSCO)\cite{Cheong} despite of their dynamical natures,
while the stripe phase in Nd-LSCO is considered static.  
In this section, we extend the model described in the previous section 
and apply to the Nd-LSCO system. It is one of the key issues
whether the stripe phase is intrinsically metallic or insulating.
Here, however, our study will proceed based on the simple
assumption of metallic one-dimensional system.
As assumed before,
our model has completely suppressed 
the interchain (interstripe) interaction and out-of-chain excitation
and may be more suitable for describing the static stripe 
phase. So Nd-LSCO will be preferred 
in our theory. Nd-LSCO has characteristic stripe (one-dimensional) phases,
where the stripes in adjacent planes are rotated by $\frac{\pi}{2}$,
whose schematic sketch is given in Fig.\ref{fig:NdLSCO-stripe}. 
It should be noted, however, that the ARPES study of LSCO
with dynamical stripes shows similar characteristics such as 
one-dimension-like low energy electronic structure
and the suppression of nodal states\cite{Ino}. Therefore, 
the studies of static stripes would help to understand the dynamical 
stripes, too.

In the study, we are bearing La$_{1.28}$Nd$_{0.6}$Sr$_{0.12}$CuO$_4$ 
in our mind as being the reference compound,
available in the recent ARPES experiment\cite{Zhou}. In this
doping, each charged stripe is separated by three undoped stripes,
so the system has a periodicity of $4l$ 
($l$ is the lattice constant in the in-plane
direction), whereas the stripe itself is $1/4$ filled.
Then the in-plane interchain (between charged stripes) distance 
$\cpr$ would be $\cpr=4l$ and the out-of-plane distance would be
$\cpp\sim 2l$. Here, the wave vector unit should be $\pi/l$.
In the unit, the Fermi wave vector is $1/4$.
In principle, the starting point is exactly same as in PBCO, in the previous
section III except that each stripe is assumed a Fermi liquid
rather than a Tomonaga-Luttinger liquid.
For describing the stripes perpendicularly crossed layer by layer,
we write the Green's function of the system 
\begin{eqnarray}\label{IV1}
G_i({\bf r},{\bf r}^{\prime})&=&\bar{G}_i(y-y^{\prime})
          \sum_{m=-\infty}^{\infty}\chi(x-\cpr m)\chi^{\ast}(x^{\prime}-\cpr m)
\\ \nonumber &\times&
          \sum_{m^{\prime}=0}^{\infty}
               \chi(z-2\cpp m^{\prime}-\frac{\cpp}{2})
               \chi^{\ast}(z^{\prime}-2\cpp m^{\prime}-\frac{\cpp}{2})
\\ \nonumber
          &+&\bar{G}_i(x-x^{\prime})
          \sum_{m=-\infty}^{\infty}\chi(y-\cpr m)\chi^{\ast}(y^{\prime}-\cpr m)
\\ \nonumber &\times&
          \sum_{m^{\prime}=0}^{\infty}
               \chi(z-2\cpp m^{\prime}-\frac{3\cpp}{2})
               \chi^{\ast}(z^{\prime}-2\cpp m^{\prime}-\frac{3\cpp}{2})
\\ \nonumber
          &=&G_{i1}({\bf r},{\bf r}^{\prime})
            +G_{i2}({\bf r},{\bf r}^{\prime}), \ \ i=1,2
\end{eqnarray}
where $i=1,2$ denote the Green function at $E$ and $E+\omega$, respectively.
The surface is assumed to be at $z=0$. The Green's function matrix element
$\langle{\bf \tilde{k}}|G_i|{\bf k}^{\prime}\rangle$ is 
\begin{eqnarray}\label{IV2}
\langle{\bf \tilde{k}}|G_i|{\bf k}^{\prime}\rangle
&=&\bar{G}_i(k_y)\delta(k_y-k_y^{\prime})
\nonumber \\ &\times&
 \chi(k_x)\chi^{\ast}(k_x^{\prime})
 \left[\frac{1}{\cpr}\sum_{\Gpr}\delta(k_x-k_x^{\prime}+\Gpr)\right]
\nonumber \\ &\times&
 \sum_{m=0}^{\infty}\tilde{\chi}^m(k_z)
            [\bar{\chi}^m(k_z^{\prime})]^{\ast}
\nonumber \\
&+&\bar{G}_i(k_x)\delta(k_x-k_x^{\prime})
\nonumber \\ &\times&
 \chi(k_y)\chi^{\ast}(k_y^{\prime})
 \left[\frac{1}{\cpr}\sum_{\Gpr}\delta(k_y-k_y^{\prime}+\Gpr)\right]
\nonumber \\ &\times&
 \sum_{m=0}^{\infty}\tilde{\chi}^{m+\frac{1}{2}}(k_z)
            [\bar{\chi}^{m+\frac{1}{2}}(k_z^{\prime})]^{\ast}
\nonumber \\
&=&\langle{\bf \tilde{k}}|G_{i1}|{\bf k}^{\prime}\rangle
 +\langle{\bf \tilde{k}}|G_{i2}|{\bf k}^{\prime}\rangle,
\end{eqnarray}
where $\Gpr=2n\pi/\cpr$ is the reciprocal lattice vector 
for an in-plane direction.
Following the same line of approximation as in the previous PBCO case,
we note 
$$
\bar{\chi}^m(k_z)=\frac{1}{\sqrt{2\pi a}}\frac{2/a}{1/a^2+k_z^2}
                  e^{ik_z(2\cpp m+\cpp/2)},
$$
\bb\label{IV3}
\tilde{\chi}^m(k_z)=\frac{1}{\sqrt{2\pi a}}\frac{2/a}{1/a^2+k_z^2}
                  e^{i\tilde{k}_z(2\cpp m+\cpp/2)}.
\ee
Then we see the photoemission intensity $I({\bf k},E,\omega)$ 
is expressed as (adopting the simplified notation),
for $\Delta=\widehat{E}\cdot{\bf p}$, 
\begin{eqnarray}\label{IV4}
I({\bf k},E,\omega)&\propto&G^{+}_{21}\Delta G^{+}_{11}\Delta G^{-}_{21}
                   +G^{+}_{22}\Delta G^{+}_{12}\Delta G^{-}_{22}
\nonumber \\ 
             &+&G^{+}_{22}\Delta G^{+}_{11}\Delta G^{-}_{22}
              +G^{+}_{21}\Delta G^{+}_{12}\Delta G^{-}_{21}
\nonumber \\
             &+&G^{+}_{21}\Delta G^{+}_{11}\Delta G^{-}_{22}
              +G^{+}_{22}\Delta G^{+}_{11}\Delta G^{-}_{21}
\nonumber \\
             &+&G^{+}_{21}\Delta G^{+}_{12}\Delta G^{-}_{22}
              +G^{+}_{22}\Delta G^{+}_{12}\Delta G^{-}_{21}, 
\end{eqnarray}
where we readily recognize only the first two terms 
($G^{+}_{21}\Delta G^{+}_{11}\Delta G^{-}_{21}$ and 
$G^{+}_{22}\Delta G^{+}_{12}\Delta G^{-}_{22}$) are the direct
term for each $y-$ and $x-$directed chain set with
the perpendicular interchain distance $2\cpp$, while 
the others are given by the complex interferences of two perpendicular sets.
The calculation can be done in the same way as in the section III,
but its algebra is quite long and tedious. The final expression 
has a slight asymmetry in interchanging $x-$ and $y-$coordinates
depending on which set is the first layer. Thanks to the remaining
symmetries in Eq.(\ref{IV4}), nevertheless,
it will be enough to calculate the half of terms
and the other half could be obtained by interchanging
$k_x\leftrightarrow k_y$ and $\varepsilon_x\leftrightarrow\varepsilon_y$
if only one account for some damping terms.

In Eq.(\ref{IV4}), $G^{+}_{21}\Delta G^{+}_{11}\Delta G^{-}_{21}$ is
exactly same calculation as in Eq.(\ref{III19}) except for a little
changes like $\cpp\to 2\cpp$ in $\alpha(k_z)$, $\beta(k_z)$,
and $\gamma(k_z)$. Here for a later use, we redefine $\alpha(k_z)$, 
$\beta(k_z)$, $\gamma(k_z)$, $\bar{\xi}(m)$, and $\bar{\zeta}(m)$
by putting $\cpp\to 2\cpp$. 
For $G^{+}_{22}\Delta G^{+}_{11}\Delta G^{-}_{22}$ in Eq.(\ref{IV4}), 
we have the expression, apart from irrelevant constants, 
\begin{eqnarray}\label{IV5}
& &G^{+}_{22}\Delta G^{+}_{11}\Delta G^{-}_{22}
\\ \nonumber
&=&\left[-\frac{1}{\pi}{\rm Im}\bar{G}_1^+(k_y)\right]
   |\chi(k_x)|^2|\chi(k_y)|^2|\chi(k_z)|^2
   e^{-3{\rm Im}\tilde{k}_z \cpp}
\\ \nonumber
&\times& 
   \sum_m\sum_{m^{\prime}}\sum_{m^{\prime\prime}}
   e^{-2{\rm Im}\tilde{k}_z \cpp(m+m^{\prime\prime})}
   e^{i2k_z \cpp(m-m^{\prime\prime})}
\\ \nonumber
&\times&
   [\varepsilon_x^2k_x^2\xi(k_y)\bar{\xi}(m-m^{\prime}+\frac{1}{2})
    \bar{\xi}(m^{\prime}-m^{\prime\prime}-\frac{1}{2})
\\ \nonumber
& & 
   +\varepsilon_y^2\eta(k_y)\bar{\xi}(m-m^{\prime}+\frac{1}{2})
    \bar{\xi}(m^{\prime}-m^{\prime\prime}-\frac{1}{2})
\\ \nonumber
& & 
   +\varepsilon_z^2\xi(k_y)\bar{\zeta}(m-m^{\prime}+\frac{1}{2})
     \bar{\zeta}(m^{\prime}-m^{\prime\prime}-\frac{1}{2})
\\ \nonumber
& & 
   +2\varepsilon_x\varepsilon_yk_x\zeta(k_y)\bar{\xi}(m-m^{\prime}+\frac{1}{2})
     \bar{\xi}(m^{\prime}-m^{\prime\prime}-\frac{1}{2})
\\ \nonumber
& & 
   +\varepsilon_y\varepsilon_z\zeta(k_y)\bar{\xi}(m-m^{\prime}+\frac{1}{2})
     \bar{\zeta}(m^{\prime}-m^{\prime\prime}-\frac{1}{2})
\\ \nonumber
& & 
   +\varepsilon_y\varepsilon_z\zeta(k_y)\bar{\zeta}(m-m^{\prime}+\frac{1}{2})
     \bar{\xi}(m^{\prime}-m^{\prime\prime}-\frac{1}{2})
\\ \nonumber
& & 
   +\varepsilon_z\varepsilon_xk_x\xi(k_y)\bar{\zeta}(m-m^{\prime}+\frac{1}{2})
     \bar{\xi}(m^{\prime}-m^{\prime\prime}-\frac{1}{2})
\\ \nonumber
& &
   +\varepsilon_z\varepsilon_xk_x\xi(k_y)\bar{\xi}(m-m^{\prime}+\frac{1}{2})
     \bar{\zeta}(m^{\prime}-m^{\prime\prime}-\frac{1}{2})],
\end{eqnarray}
where $\eta(k)$ is another $\Gpr$-summation 
not appeared in section III,
\begin{eqnarray}\label{IV6}
& &\sum_{\Gpr}(k+\Gpr)^2|\chi(k+\Gpr)|^2
\nonumber \\
&=&\frac{1}{2\pi a}\frac{\cpr}{a}
   \sum_m e^{ik\cpr m}[-\cpr|m|/a+1]e^{-\cpr|m|/a}
\nonumber \\
    &\approx&\frac{1}{2\pi a}\frac{\cpr}{a}
    [1-2\cos(k\cpr)(\cpr/a-1)e^{-\cpr/a}
\nonumber \\& & \ \ \ \ \ \ \ \ \ \ \ \
      -2\cos(2k\cpr)(2\cpr/a-1)e^{-2\cpr/a}]
\nonumber \\
    &\equiv&\eta(k),
\end{eqnarray}
and the other sophisticated summations for $m$, $m^{\prime}$, 
and $m^{\prime\prime}$ (all $m$'s $>0$) can be managed by considering the small
parameter $e^{-\cpp/a}$,
\begin{eqnarray}\label{IV7}
\bar{\alpha}(k_z)&\equiv&\sum_m\sum_{m^{\prime}}\sum_{m^{\prime\prime}}
    e^{-2{\rm Im}\tilde{k}_z\cpp(m+m^{\prime\prime})}
    e^{i2k_z\cpp(m-m^{\prime\prime})}
\nonumber \\ &\times&
    \bar{\xi}(m-m^{\prime}\pm \frac{1}{2})
    \bar{\xi}(m^{\prime}-m^{\prime\prime}\mp \frac{1}{2})
\nonumber \\
    &\approx&
    2(\cpp/a+1)^2[1+\cos(2k_z\cpp)e^{-2{\rm Im}\tilde{k}_z \cpp}]e^{-2\cpp/a}
\nonumber \\ &\times&
    \sum_{m=0}^{\infty}e^{-4{\rm Im}\tilde{k}_z \cpp m},
\end{eqnarray}
\begin{eqnarray}\label{IV8}
\bar{\beta}(k_z)&\equiv&
    \sum_m\sum_{m^{\prime}}\sum_{m^{\prime\prime}}
    e^{-2{\rm Im}\tilde{k}_z\cpp(m+m^{\prime\prime})}
    e^{i2k_z\cpp(m-m^{\prime\prime})}
\nonumber \\ &\times&
    \bar{\zeta}(m-m^{\prime}\pm \frac{1}{2})
    \bar{\zeta}(m^{\prime}-m^{\prime\prime}\mp \frac{1}{2})
\nonumber \\
    &\approx&
    \frac{2}{a^2}\frac{\cpp^2}{a^2}[1-\cos(2k_z\cpp)
    e^{-2{\rm Im}\tilde{k}_z \cpp}]e^{-2\cpp/a}
\nonumber \\ &\times&
    \sum_{m=0}^{\infty}e^{-4{\rm Im}\tilde{k}_z \cpp m},
\end{eqnarray}
\begin{eqnarray}\label{IV9}
\bar{\gamma}(k_z)&\equiv&
    \sum_m\sum_{m^{\prime}}\sum_{m^{\prime\prime}}
    e^{-2{\rm Im}\tilde{k}_z\cpp(m+m^{\prime\prime})}
    e^{i2k_z\cpp(m-m^{\prime\prime})}
\nonumber \\ &\times&
    \bar{\xi}(m-m^{\prime}\pm \frac{1}{2})
    \bar{\zeta}(m^{\prime}-m^{\prime\prime}\mp \frac{1}{2})
\\ \nonumber
    &=&\sum_m\sum_{m^{\prime}}\sum_{m^{\prime\prime}}
    e^{-2{\rm Im}\tilde{k}\cpp(m+m^{\prime\prime})}
    e^{i2k_z\cpp(m-m^{\prime\prime})}
\nonumber \\ &\times&
    \bar{\zeta}(m-m^{\prime}\pm \frac{1}{2})
    \bar{\xi}(m^{\prime}-m^{\prime\prime}\mp \frac{1}{2})
\\ \nonumber
    &\approx&
    \frac{2}{a}\sin(2k_z \cpp)(\cpp/a)(\cpp/a+1)
    e^{-2\cpp/a}e^{-2{\rm Im}\tilde{k}_z\cpp}
\nonumber \\ &\times&
    \sum_{m=0}^{\infty}e^{-4{\rm Im}\tilde{k}_z \cpp m}.
\end{eqnarray}
The calculated $\bar{\alpha}(k_z)$, $\bar{\beta}(k_z)$, and 
$\bar{\gamma}(k_z)$ are correct to up to $e^{-3\cpp/a}$.
Then $G^{+}_{22}\Delta G^{+}_{11}\Delta G^{-}_{22}$ is written down
\begin{eqnarray}\label{IV10}
& &G^{+}_{22}\Delta G^{+}_{11}\Delta G^{-}_{22}
\nonumber \\
&=&\left[-\frac{1}{\pi}{\rm Im}\bar{G}_1^+(k_y)\right]
  |\chi(k_x)|^2|\chi(k_y)|^2|\chi(k_z)|^2
   e^{-3{\rm Im}\tilde{k}_z c}
\nonumber \\
  &\times&
  [\varepsilon_x^2k_x^2\xi(k_y)\bar{\alpha}(k_z)
  +\varepsilon_y^2\eta(k_y)\bar{\alpha}(k_z)
\nonumber \\& &
  +\varepsilon_z^2\xi(k_y)\bar{\beta}(k_z)
  +2\varepsilon_x\varepsilon_yk_x\zeta(k_y)\bar{\alpha}(k_z)
\nonumber \\& &
  +2\varepsilon_y\varepsilon_z\zeta(k_y)\bar{\gamma}(k_z)
  +2\varepsilon_z\varepsilon_xk_x\xi(k_y)\bar{\gamma}(k_z)].
\end{eqnarray}
In the calculations of $G^{+}_{21}\Delta G^{+}_{11}\Delta G^{-}_{21}$
and $G^{+}_{22}\Delta G^{+}_{11}\Delta G^{-}_{22}$, we have neglected 
$|\bar{G}_2(k_y)|^2$ (for the latter, $|\bar{G}_2(k_x)|^2$) 
based on assuming that the final state distribution
may be structureless and smooth from $E+\omega\gg E$. 
Next we go to the evaluation of
$G^{+}_{21}\Delta G^{+}_{11}\Delta G^{-}_{22}
+G^{+}_{22}\Delta G^{+}_{11}\Delta G^{-}_{21}$. Here we need 
a more drastic assumption on behaviors of $\bar{G}_2(k_y)$, i.e.,  
${\rm Im}\bar{G}_2(k_y)\approx 0$. Thus we assume 
$\bar{G}_2^+(k_y)\bar{G}_2^-(k_x)$ or 
$\bar{G}_2^+(k_x)\bar{G}_2^-(k_y)$ could be also taken just as 
a positive constant and further
$|\bar{G}_2(k_y)|^2\approx|\bar{G}_2(k_x)|^2\approx
\bar{G}_2^+(k_y)\bar{G}_2^-(k_x)\approx\bar{G}_2^+(k_x)\bar{G}_2^-(k_y)$.
Then $G^{+}_{21}\Delta G^{+}_{11}\Delta G^{-}_{22}$ is 
\begin{eqnarray}\label{IV11}
& &G^{+}_{21}\Delta G^{+}_{11}\Delta G^{-}_{22}
\nonumber \\
&=&\left[-\frac{1}{\pi}{\rm Im}\bar{G}_1^+(k_y)\right]
   |\chi(k_x)|^2|\chi(k_y)|^2|\chi(k_z)|^2
   e^{-2{\rm Im}\tilde{k}_z \cpp}
\nonumber \\ &\times&
   \sum_m\sum_{m^{\prime}}\sum_{m^{\prime\prime}}
   e^{-2{\rm Im}\tilde{k}_z \cpp(m+m^{\prime\prime})}
   e^{i2k_z\cpp[m-m^{\prime\prime}-\frac{1}{2}]}
\nonumber \\
&\times&
   [\varepsilon_x^2 k_x\zeta(k_x)\bar{\xi}(m-m^{\prime})
    \bar{\xi}(m^{\prime}-m^{\prime\prime}-\frac{1}{2})
\nonumber \\
& &
   +\varepsilon_y^2 k_y^2\xi(k_x)\bar{\xi}(m-m^{\prime})
    \bar{\xi}(m^{\prime}-m^{\prime\prime}-\frac{1}{2})
\nonumber \\
& &
   +\varepsilon_z^2\xi(k_x)\bar{\zeta}(m-m^{\prime})
    \bar{\zeta}(m^{\prime}-m^{\prime\prime}-\frac{1}{2})
\nonumber \\
& &
   +\varepsilon_x\varepsilon_yk_y\zeta(k_x)\bar{\xi}(m-m^{\prime})
    \bar{\xi}(m^{\prime}-m^{\prime\prime}-\frac{1}{2})
\nonumber \\
& &
   +\varepsilon_x\varepsilon_yk_xk_y\xi(k_x)\bar{\xi}(m-m^{\prime})
    \bar{\xi}(m^{\prime}-m^{\prime\prime}-\frac{1}{2})
\nonumber \\
& &
   +\varepsilon_y\varepsilon_zk_y\xi(k_x)\bar{\xi}(m-m^{\prime})
    \bar{\zeta}(m^{\prime}-m^{\prime\prime}-\frac{1}{2})
\nonumber \\
& &
   +\varepsilon_y\varepsilon_zk_y\xi(k_x)\bar{\zeta}(m-m^{\prime})
    \bar{\xi}(m^{\prime}-m^{\prime\prime}-\frac{1}{2})
\nonumber \\
& &
   +\varepsilon_z\varepsilon_xk_x\xi(k_x)\bar{\zeta}(m-m^{\prime})
    \bar{\xi}(m^{\prime}-m^{\prime\prime}-\frac{1}{2})
\nonumber \\
& & 
   +\varepsilon_z\varepsilon_x\zeta(k_x)\bar{\xi}(m-m^{\prime})
    \bar{\zeta}(m^{\prime}-m^{\prime\prime}-\frac{1}{2})],
\end{eqnarray}
and $G^{+}_{22}\Delta G^{+}_{11}\Delta G^{-}_{21}$ is also similarly
\begin{eqnarray}\label{IV12}
& &G^{+}_{22}\Delta G^{+}_{11}\Delta G^{-}_{21}
\nonumber \\
&=&
   \left[-\frac{1}{\pi}{\rm Im}\bar{G}_1^+(k_y)\right]
   |\chi(k_x)|^2|\chi(k_y)|^2|\chi(k_z)|^2
   e^{-2{\rm Im}\tilde{k}_z \cpp} 
\nonumber \\ &\times&
   \sum_m\sum_{m^{\prime}}\sum_{m^{\prime\prime}}
   e^{-2{\rm Im}\tilde{k}_z \cpp(m+m^{\prime\prime})}
   e^{i2k_z\cpp[m-m^{\prime\prime}+\frac{1}{2}]}
\nonumber \\ &\times&
   [\varepsilon_x^2 k_x\zeta(k_x)\bar{\xi}(m-m^{\prime}+\frac{1}{2})
    \bar{\xi}(m^{\prime}-m^{\prime\prime})
\nonumber \\   
& & 
   +\varepsilon_y^2 k_y^2\xi(k_x)\bar{\xi}(m-m^{\prime}+\frac{1}{2})
    \bar{\xi}(m^{\prime}-m^{\prime\prime})
\nonumber \\
& &
   +\varepsilon_z^2\xi(k_x)\bar{\zeta}(m-m^{\prime}+\frac{1}{2})
    \bar{\zeta}(m^{\prime}-m^{\prime\prime})
\nonumber \\   
& & 
   +\varepsilon_x\varepsilon_yk_y\zeta(k_x)\bar{\xi}(m-m^{\prime}+\frac{1}{2})
    \bar{\xi}(m^{\prime}-m^{\prime\prime})
\nonumber \\
& &
   +\varepsilon_x\varepsilon_yk_xk_y\xi(k_x)\bar{\xi}(m-m^{\prime}+\frac{1}{2})
    \bar{\xi}(m^{\prime}-m^{\prime\prime})
\nonumber \\   
& & 
   +\varepsilon_y\varepsilon_zk_y\xi(k_x)\bar{\xi}(m-m^{\prime}+\frac{1}{2})
    \bar{\zeta}(m^{\prime}-m^{\prime\prime})
\nonumber \\
& &
   +\varepsilon_y\varepsilon_zk_y\xi(k_x)\bar{\zeta}(m-m^{\prime}+\frac{1}{2})
    \bar{\xi}(m^{\prime}-m^{\prime\prime})
\nonumber \\   
& & 
   +\varepsilon_z\varepsilon_xk_x\xi(k_x)\bar{\xi}(m-m^{\prime}+\frac{1}{2})
    \bar{\zeta}(m^{\prime}-m^{\prime\prime})
\nonumber \\
& & 
   +\varepsilon_z\varepsilon_x\zeta(k_x)\bar{\zeta}(m-m^{\prime}+\frac{1}{2})
    \bar{\xi}(m^{\prime}-m^{\prime\prime})].
\end{eqnarray}
We can also define $\tilde{\alpha}(k_z)$, $\tilde{\beta}(k_z)$,
$\tilde{\gamma}_1(k_z)$, and $\tilde{\gamma}_1(k_z)$ as the 
summations for $m$'s relevant in Eqs.(\ref{IV11}) and (\ref{IV12}); 
\begin{eqnarray}\label{IV13}
\tilde{\alpha}(k_z)&\equiv&
    {\rm Re}\sum_m\sum_{m^{\prime}}\sum_{m^{\prime\prime}}
    e^{-2{\rm Im}\tilde{k}_z\cpp(m+m^{\prime\prime})}
    e^{i2k_z\cpp(m-m^{\prime\prime}-\frac{1}{2})}
\nonumber \\ &\times&
     \bar{\xi}(m-m^{\prime})\bar{\xi}(m^{\prime}-m^{\prime\prime}-\frac{1}{2})
\nonumber \\
    &\approx&\cos(k_z\cpp)(\cpp/a+1)[1+(2\cpp/a+1)e^{-2\cpp/a}]
\nonumber \\ &\times&
      e^{-\cpp/a}(1+e^{-2{\rm Im}\tilde{k}_z\cpp})
      \sum_{m=0}^{\infty}e^{-4{\rm Im}\tilde{k}_z \cpp m}
\nonumber \\
    &+&\cos(3k_z\cpp)(2\cpp/a+1)(\cpp/a+1)
      e^{-2{\rm Im}\tilde{k}_z\cpp}
\nonumber \\ &\times&
      e^{-3\cpp/a}(1+e^{-2{\rm Im}\tilde{k}_z\cpp})
      \sum_{m=0}^{\infty}e^{-4{\rm Im}\tilde{k}_z\cpp m},
\end{eqnarray} 
\begin{eqnarray}\label{IV14}
\tilde{\beta}(k_z)&\equiv&
    {\rm Re}\sum_m\sum_{m^{\prime}}\sum_{m^{\prime\prime}}
    e^{-2{\rm Im}\tilde{k}_z\cpp(m+m^{\prime\prime})}
    e^{i2k_z\cpp(m-m^{\prime\prime}-\frac{1}{2})}
\nonumber \\ &\times&
    \bar{\zeta}(m-m^{\prime})
    \bar{\zeta}(m^{\prime}-m^{\prime\prime}-\frac{1}{2})
\nonumber \\
    &\approx&\frac{2}{a^2}\frac{\cpp^2}{a^2}
    [\cos(k_z \cpp)-\cos(3k_z \cpp)e^{-2{\rm Im}\tilde{k}_z\cpp}]
\nonumber \\ &\times&
    e^{-3\cpp/a}(1+e^{-2{\rm Im}\tilde{k}_z\cpp})
    \sum_{m=0}^{\infty}e^{-4{\rm Im}\tilde{k}_z \cpp m},
\end{eqnarray}
\begin{eqnarray}\label{IV15}
\tilde{\gamma}_1(k_z)&\equiv&
    {\rm Re}\sum_m\sum_{m^{\prime}}\sum_{m^{\prime\prime}}
    e^{-2{\rm Im}\tilde{k}_z\cpp(m+m^{\prime\prime})}
    e^{i2k_z\cpp(m-m^{\prime\prime}-\frac{1}{2})}
\nonumber \\ &\times&
    \bar{\xi}(m-m^{\prime})
    \bar{\zeta}(m^{\prime}-m^{\prime\prime}-\frac{1}{2})
\nonumber \\
    &\approx&
    \frac{1}{a}\frac{\cpp}{a}\sin(k_z\cpp)e^{-\cpp/a}[1-(2\cpp/a+1)e^{-2\cpp/a}]
\nonumber \\ &\times&
    (1+e^{-2{\rm Im}\tilde{k}_z\cpp})
    \sum_{m=0}^{\infty}e^{-4{\rm Im}\tilde{k}_z \cpp m}
\nonumber \\
    &+&\frac{1}{a}\frac{\cpp}{a}\sin(3k_z \cpp)(2\cpp/a+1)e^{-3\cpp/a}
    e^{-2{\rm Im}\tilde{k}_z\cpp}
\nonumber \\ &\times&
    (1+e^{-2{\rm Im}\tilde{k}_z\cpp})
    \sum_{m=0}^{\infty}e^{-4{\rm Im}\tilde{k}_z \cpp m},
\end{eqnarray}
\begin{eqnarray}\label{IV16}
\tilde{\gamma}_2(k_z)&\equiv&
    {\rm Re}\sum_m\sum_{m^{\prime}}\sum_{m^{\prime\prime}}
    e^{-2{\rm Im}\tilde{k}_z\cpp(m+m^{\prime\prime})}
    e^{i2k_z\cpp(m-m^{\prime\prime}-\frac{1}{2})}
\nonumber \\ &\times&
    \bar{\zeta}(m-m^{\prime})
    \bar{\xi}(m^{\prime}-m^{\prime\prime}-\frac{1}{2})
\nonumber \\
    &\approx&\frac{2}{a}\frac{\cpp}{a}
    [\sin(k_z \cpp)+\sin(3k_z\cpp)e^{-2{\rm Im}\tilde{k}_z\cpp}](\cpp/a+1)
\nonumber \\ &\times&
    e^{-3\cpp/a}(1+e^{-2{\rm Im}\tilde{k}_z\cpp})
    \sum_{m=0}^{\infty}e^{-4{\rm Im}\tilde{k}_z \cpp m}.
\end{eqnarray}
The above $\tilde{\alpha}(k_z)$, $\tilde{\beta}(k_z)$,
$\tilde{\gamma}_1(k_z)$, and $\tilde{\gamma}_1(k_z)$ are also
correct up to $e^{-3\cpp/a}$. Now we have done the evaluation of
$G^{+}_{21}\Delta G^{+}_{11}\Delta G^{-}_{22}
+G^{+}_{22}\Delta G^{+}_{11}\Delta G^{-}_{21}$. 
Using Eqs.(\ref{IV13})-(\ref{IV16}), we have
\begin{eqnarray}\label{IV17}
& &G^{+}_{21}\Delta G^{+}_{11}\Delta G^{-}_{22}
+G^{+}_{22}\Delta G^{+}_{11}\Delta G^{-}_{21}
\nonumber \\
&=&2\left[-\frac{1}{\pi}{\rm Im}\bar{G}_1^+(k_y)\right]
   |\chi(k_x)|^2|\chi(k_y)|^2|\chi(k_z)|^2
   e^{-2{\rm Im}\tilde{k}_z \cpp}
\nonumber \\
&\times&
   [\varepsilon_x^2 k_x\zeta(k_x)\tilde{\alpha}(k_z)
   +\varepsilon_y^2 k_y^2\xi(k_x)\tilde{\alpha}(k_z)
\nonumber \\ & &
   +\varepsilon_z^2\xi(k_x)\tilde{\beta}(k_z)
   +\varepsilon_x\varepsilon_yk_y\zeta(k_x)\tilde{\alpha}(k_z)
\nonumber \\ & &
   +\varepsilon_x\varepsilon_yk_xk_y\xi(k_x)\tilde{\alpha}(k_z)
   +\varepsilon_y\varepsilon_zk_y\xi(k_x)\tilde{\gamma}_1(k_z)
\nonumber \\ & &
   +\varepsilon_y\varepsilon_zk_y\xi(k_x)\tilde{\gamma}_2(k_z)
   +\varepsilon_z\varepsilon_xk_x\xi(k_x)\tilde{\gamma}_2(k_z)
\nonumber \\ & &
   +\varepsilon_z\varepsilon_x\zeta(k_x)\tilde{\gamma}_1(k_z)].
\end{eqnarray}
The other four terms out of Eq.(\ref{IV4})
can be obtained using the symmetry above without actual calculations.
From Eq.(\ref{IV4}), we note the other terms are 
\begin{eqnarray}\label{IV18}
G^{+}_{22}\Delta G^{+}_{12}\Delta G^{-}_{22}
&=&e^{-2{\rm Im}\tilde{k}_z \cpp}
\nonumber  \\
&\times&
 [G^{+}_{21}\Delta G^{+}_{11}\Delta G^{-}_{21}; \
 x \Longleftrightarrow y],
\end{eqnarray}
where the notation $[G^{+}_{21}\Delta G^{+}_{11}\Delta G^{-}_{21}; \
x \Longleftrightarrow y]$
means the same expression as $G^{+}_{21}\Delta G^{+}_{11}\Delta G^{-}_{21}$
with only $k_x\leftrightarrow k_y$, $\varepsilon_x\leftrightarrow\varepsilon_y$
exchanged. When keeping the same notation,
\begin{eqnarray}\label{IV19}
G^{+}_{21}\Delta G^{+}_{12}\Delta G^{-}_{21}
&=&e^{2{\rm Im}\tilde{k}_z \cpp}
\nonumber  \\
&\times&
 [G^{+}_{22}\Delta G^{+}_{11}\Delta G^{-}_{22}; \
 x \Longleftrightarrow y],
\end{eqnarray}
\begin{eqnarray}\label{IV20}
& &G^{+}_{21}\Delta G^{+}_{12}\Delta G^{-}_{22}
+G^{+}_{22}\Delta G^{+}_{12}\Delta G^{-}_{21}
\nonumber \\
&=&[G^{+}_{21}\Delta G^{+}_{11}\Delta G^{-}_{22}
 +G^{+}_{22}\Delta G^{+}_{11}\Delta G^{-}_{21}; \ x \Longleftrightarrow y].
\end{eqnarray}
The total photoemission intensity $I({\bf k},E,\omega)$ is the sum of
permutative contributions (Eq.(\ref{IV4})) obtained above.

In the experiment of Zhou {\it et al.}'s\cite{Zhou}, 
the spectral weight distribution has been 
represented by integrating $I({\bf k},E,\omega)$ for finite 
energy windows. From the weight distribution, they
suggest the superposition of two perpendicular one-dimensional 
charged stripes for the underlying electronic structure.
Other features such as the weight suppression along the  
$d$-wave nodal directions ${\bf k}=(\pm 1,\pm 1)$ 
including the $\Gamma$ point does not seem easily explained,
for which they propose a gap formation 
around $|k_x|\sim 1/4$ and $|k_y|\sim 1/4$. The origin of the gap,
however, does not look very clear. Thus there may be left some rooms for
another possibility. In our calculation, 
the integration of $I({\bf k},E,\omega)$ with respect to $E$ 
can be readily done. From Eqs.(\ref{IV18}), (\ref{IV19}), and (\ref{IV20}),
the $E$-dependency is in 
$\left[-\frac{1}{\pi}{\rm Im}\bar{G}_1^+(k_y)\right]$ or
$\left[-\frac{1}{\pi}{\rm Im}\bar{G}_1^+(k_x)\right]$
and through $k_z$. For $k_z=\sqrt{2(E+\omega)-k_x^2-k_y^2}$, 
$|E|\lesssim 0.1$ eV from the band width, while $\omega\sim{\cal O}(10)$ eV 
in actual experiments. So we could approximate 
$k_z\approx\sqrt{2\omega-k_x^2-k_y^2}$. Then the remaining integral is 
special,
\bb\label{IV21}
\int dE\left[-\frac{1}{\pi}{\rm Im}\bar{G}_1^+(k_{y(x)})f(E)\right]
=n_{k_{y(x)}},
\ee
where $f(E)$ is the Fermi-distribution function.
$\left[-\frac{1}{\pi}{\rm Im}\bar{G}_1^+(k_{y(x)})\right]$ is the
spectral function of the stripe, which determines  
its band structure. Because our starting point is 
the noninteracting chain (stripe), we should assume a suitable 
band dispersion relevant to the one-dimensional stripe.
The corresponding band by the model calculation\cite{Salkola}
from vertical stripes is very flat especially near $(\pi,0)$, which
is believed to have the one-dimensional character in the complicated 
real band dispersion. In the experiments\cite{Zhou00}, its band width
is estimated about 30 meV and should be the integration energy window
in our study. So we assume the band dispersion of
the stripe to be $b k_y^2-b k_f^2$ for $\hat{y}$-directed stripe,
where $b\sim 0.18$ from $(\pi/l)^2\sim 2.64$ eV ($l\sim 5.34$\AA).
It should be noticed that, 
despite one-dimensional properties in the Nd-LSCO system,
there have been no compelling evidences for Tomonaga-Luttinger-like
electronic structure (i.e., such as the spin-charge separation) 
in the stripe phase. Therefore, even if the matrix effects are
explored based on one-dimensional stripes, we assume the Fermi-like
momentum distribution $n_{k_{y(x)}}$, which is motivated by our interests
in the polarization, photon energy and momentum dependencies 
of the ARPES intensity, not in the energy distribution curves (EDC's).  
Then we get the expression $\bar{I}({\bf k},\omega)$,
defining $\bar{I}({\bf k},\omega)\equiv\int dE I({\bf k},E,\omega)f(E)$,
\begin{eqnarray}\label{IV22}
\bar{I}({\bf k},\omega)&\propto&n_{k_y}
       |\chi(k_x)|^2|\chi(k_z)|^2e^{-{\rm Im}\tilde{k}_z \cpp}
\nonumber \\
  &\times&
  [\varepsilon_x^2\zeta(k_x)^2\alpha(k_z)
  +\varepsilon_y^2k_y^2\xi(k_x)^2\alpha(k_z)
\nonumber \\ & &
  +\varepsilon_z^2\xi(k_x)^2\beta(k_z)
  +2\varepsilon_x\varepsilon_y k_y\xi(k_x)\zeta(k_x)\alpha(k_z)
\nonumber \\ & &
  +2\varepsilon_y\varepsilon_zk_y\xi(k_x)^2 \gamma(k_z)
  +2\varepsilon_z\varepsilon_x\xi(k_x)\zeta(k_x)\gamma(k_z)]
\nonumber \\ 
&+&e^{-2{\rm Im}\tilde{k}_z c}\times
   [k_x \Longleftrightarrow k_y,
    \varepsilon_x \Longleftrightarrow \varepsilon_y]
\nonumber \\ 
&+&2n_{k_y}|\chi(k_x)|^2|\chi(k_y)|^2|\chi(k_z)|^2
   e^{-2{\rm Im}\tilde{k}_z \cpp}
\nonumber \\ 
&\times&
   [\varepsilon_x^2 k_x\zeta(k_x)\tilde{\alpha}(k_z)
   +\varepsilon_y^2 k_y^2\xi(k_x)\tilde{\alpha}(k_z)
\nonumber \\ & &
   +\varepsilon_z^2\xi(k_x)\tilde{\beta}(k_z)
   +\varepsilon_x\varepsilon_yk_y\zeta(k_x)\tilde{\alpha}(k_z)
\nonumber \\ & &
   +\varepsilon_x\varepsilon_yk_xk_y\xi(k_x)\tilde{\alpha}(k_z)
   +\varepsilon_y\varepsilon_zk_y\xi(k_x)\tilde{\gamma}_1(k_z)
\nonumber \\ & &
   +\varepsilon_y\varepsilon_zk_y\xi(k_x)\tilde{\gamma}_2(k_z)
   +\varepsilon_z\varepsilon_xk_x\xi(k_x)\tilde{\gamma}_2(k_z)
\nonumber \\ & &
   +\varepsilon_z\varepsilon_x\zeta(k_x)\tilde{\gamma}_1(k_z)]
\nonumber \\
&+&[k_x \Longleftrightarrow k_y,
    \varepsilon_x \Longleftrightarrow \varepsilon_y]
\nonumber \\
&+&n_{k_y}|\chi(k_x)|^2|\chi(k_y)|^2|\chi(k_z)|^2
   e^{-3{\rm Im}\tilde{k}_z \cpp}
\nonumber \\ &\times&
  [\varepsilon_x^2k_x^2\xi(k_y)\bar{\alpha}(k_z)
  +\varepsilon_y^2\eta(k_y)\bar{\alpha}(k_z)
\nonumber \\ & &
  +\varepsilon_z^2\xi(k_y)\bar{\beta}(k_z)
  +2\varepsilon_x\varepsilon_yk_x\zeta(k_y)\bar{\alpha}(k_z)
\nonumber \\ & &
  +2\varepsilon_y\varepsilon_z\zeta(k_y)\bar{\gamma}(k_z)
  +2\varepsilon_z\varepsilon_xk_x\xi(k_y)\bar{\gamma}(k_z)]
\nonumber \\
  &+&e^{2{\rm Im}\tilde{k}_z \cpp}\times [k_x \Longleftrightarrow k_y,
    \varepsilon_x \Longleftrightarrow \varepsilon_y],
\end{eqnarray}
where $[k_x \Longleftrightarrow k_y,
\varepsilon_x \Longleftrightarrow \varepsilon_y]$
denotes the term right before it with $k_x \leftrightarrow k_y$ and 
$\varepsilon_x \leftrightarrow \varepsilon_y$ exchanged. 
According to Eq.(\ref{IV22}), because of $\Gamma$ (or ${\rm Im}\tilde{k}_z$),
we expect to get an asymmetry
in the weight distribution (with respect to $k_x$ and $k_y$)
depending on which is the first layer
because the mean free path could govern only the first few layers in
realistic situations.
To compare with experiment\cite{Zhou}, 
we average the spectra with respect to the stripe direction, i.e.
for the two configurations in Fig.\ref{fig:NdLSCO-stripe}.

\begin{figure}
\vspace*{9.0cm}
\includegraphics{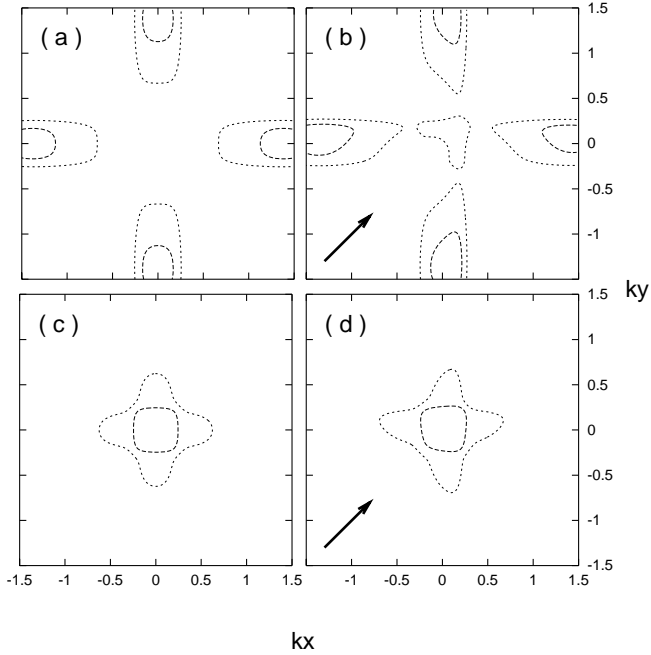}
\caption{Spectral weight distributions for the linearly polarized photon,
especially perpendicular ((a) and (c)) or nearly perpendicular ((b) and (d))
to the layer are given.
We have two kinds of distribution depending
on the photon energy. (a) and (b) are for $\omega=15(\pi/l)^2$
and (c) and (d) are for $\omega=20(\pi/l)^2$. The arrow in
the figures ((b) and (d)) denotes the direction of the in-plane electric field.
$l/a=2.5$ is used. The wave vector unit is $\pi/l$.
}
\label{fig:PEz}
\end{figure}

\begin{figure}
\vspace*{5.6cm}
\includegraphics{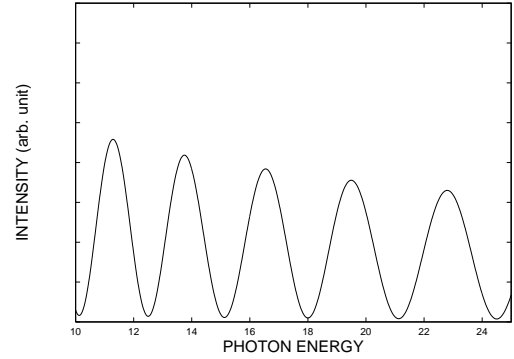}
\caption{Spectral intensities at the $\Gamma$ point are given with
respect to the photon energy when $\widehat{\bf E}\perp xy$.
The energy unit is $(\pi/l)^2$.
}
\label{fig:Gammapoint}
\end{figure}

From Eq.(\ref{IV22}), we evaluate the spectral weight distribution 
for $\widehat{\bf E}\perp xy$
and provide the results in Fig.\ref{fig:PEz}. 
Note the momentum distribution
$n_{k_{y(x)}}$ readily gives two sets of Fermi surfaces
defined by $|k_x|=1/4$ and $|k_y|=1/4$.
In the figure, we have found 
two kinds of distributions depending on the photon energy. 
Figures \ref{fig:PEz}(c)
and \ref{fig:PEz}(d) may be understood just as the superposition of two 
perpendicular sets, whereas Figs.\ref{fig:PEz}(a) and \ref{fig:PEz}(b)
are very interesting. As in the experiment,
we can observe the spectral weight reduction along
the $d$-wave nodal direction and also around the $\Gamma$ point,
and simultaneously the spectral confinement around $(0,\pm 1)$ and $(\pm 1,0)$.
This must be the geometric (interference) effect raised by the 
dipole matrix in our study. 
More clearly, when $\varepsilon_x=\varepsilon_y=0$,
$\bar{I}({\bf k},E)$ is determined by the functions $\beta(k_z)$,
$\bar{\beta}(k_z)$, and $\tilde{\beta}(k_z)$, of which 
$\bar{\beta}(k_z)$ must be dominant. Thus we can see that 
$\bar{\beta}(k_z)$ is mainly attributed to $\bar{I}({\bf k},E)$,
that is, $G^{+}_{22}\Delta G^{+}_{11}\Delta G^{-}_{22}
+G^{+}_{21}\Delta G^{+}_{12}\Delta G^{-}_{21}$,
which are the interference terms, not direct terms.
If we try to explain the experiment based on 
the present results, a gap opening is not required and metallic stripes
look enough to reproduce the observed weight distribution.
Unfortunately, however, it is also found in our study that two types of
spectral distributions appears alternatively depending on the photon energy.
Experimentally, the intensity confinement near $(0,\pm 1)$ and 
$(\pm 1,0)$ is robust when $\omega\sim20$ eV, 30 eV, and 50 eV.
In Fig.\ref{fig:Gammapoint}, we give the spectral intensities
at the $\Gamma$ point for the photon energy $\omega$. 

\begin{figure}
\vspace*{5.6cm}
\includegraphics{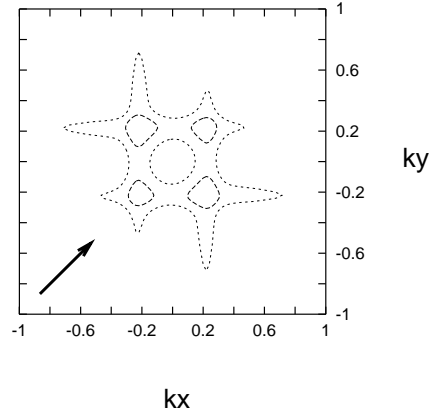}
\caption{The spectral weight distribution for
$\widehat{\bf E}\parallel xy$ for $\omega=15(\pi/l)^2$.
The arrow is the direction of $\widehat{\bf E}$.
}
\label{fig:PExy}
\end{figure}

In Fig.\ref{fig:PExy}, the spectral distribution for
$\widehat{\bf E}\parallel xy$ is given, which is quite a different distribution 
from Fig.\ref{fig:PEz}. In the figure, the intensity at the $\Gamma$ point
is suppressed and there happen to be strong weights on the diagonals, 
i.e., $(0,0)$ to $(\pm 1,\pm 1)$. According to the recent
experiment\cite{Zhou00}, for $\widehat{\bf E}$ parallel to the surface,
weak intensities appear also in the diagonal directions
in addition to the intensities at $(0,\pm 1)$ and $(\pm 1,0)$,
which would not be observed or observed weaker in the case of 
$\widehat{\bf E}$ perpendicular to the
surface. These
additional complex structures have been suggested as 
the overlap of two-dimensional characters from stripe fluctuations
on the simple one-dimensional stripe picture, namely as 
part of the two dimensional Fermi surface. On the other hand,
such fluctuations were not considered in our study.

As shown in the figures, we do not stress the restricted agreement much.
The stripe system, Nd-LSCO is a more complicated system compared to 
the chain system, PBCO. The real electronic structure may 
consist of both one-dimensional and two-dimensional components.
This is very important and can explain why the present approach is
good for PBCO, while less satisfactory for Nd-LSCO. 
The photon energy behavior in Figs.\ref{fig:PEz} and \ref{fig:Gammapoint}, 
however, provides outlook 
for further studies along the present line.
The first thing may be to go beyond the primary plane wave assumption
for the photoelectron. The photoelectron wave function
should be in principle calculated using the scattering theory 
including the surface. Also the excitation perpendicular to the stripes,
neglected here, would introduce the surface Green's function\cite{Bansil}. 
We would then hopefully expect more correct transition matrix effects,
and also more robust photon energy and polarization 
dependencies of the spectra.
However, we do not mean here the matrix effect would be enough
to understand experiments.
We have regarded stripes as a collection of noninteracting metallic chains, 
but the interaction among the chains as well as the interaction 
between the chain and the two-dimensional background
electronic structures may play a role essentially\cite{Salkola}.

\section{Conclusions}

We have studied the dipole matrix effects of the photoemission 
in the system of linear metallic chains for two characteristic 
configurations. Assuming the chains (stripes) are perfectly
one-dimensional by neglecting the transverse excitations
and interchain interactions, 
we could write down the Green's function of 
the total system. Pendrey's formula enables us to manage the 
dipole matrix effects from knowledges of the single-particle 
Green's function. The calculations are very well controlled
by small parameters $e^{-\cpr/a}$ or $e^{-\cpp/a}$.

Interesting ${\bf k}$- and $\widehat{\bf E}$-dependencies in 
the photoemission spectra in PBCO system have been observed,
which would be thought of as the transition matrix effects. 
We have evaluated the spectral intensities accounting for the
dipole matrix in the system of chains based on the Luttinger model.
PBCO is found to have one-dimensional chains 
which behave like Tomonaga-Luttinger liquid by recent ARPES studies.
For the photoemission formula, we have adopted the Pendrey's formula.
The relative intensities of separated holons and spinons
depending on the wave vector ${\bf k}$ (anisotropies), or the 
photon polarization (linear polarization, $\widehat{\bf E}$)
obtained in the study are found consistent with the recent
experiments. It is also found that the present study would be
applicable to another system having insulating Cu-O chains, SrCuO$_2$. 

We have also considered the system more extended than
PBCO. We have had interests in the static stripe phases 
observed in Nd-LSCO system. In experiments, 
their one-dimensional characters are observed from the spectral distribution
in ${\bf k}$-space, but not perfectly understood from the stripe picture alone.
We have studied its deviations arising from
the dipole matrix effects. In the same way as for PBCO,
we have calculated the integrated spectral distribution.
We then found a distribution quite consistent with experiment
(for $\widehat{\bf E}\perp xy$), 
where the intensity around the $\Gamma$ point or along the $d$-wave 
nodal directions is strongly suppressed. 
We have also found, however, another distribution showing a
strong accumulation of intensities around the $\Gamma$ point
depending on the photon energy. In the result, two types of
distributions appear alternatively as a function of photon energy.
Such more or less shaky behaviors may be due to 
the simple plane wave assumption for the photoelectron state.
For $\widehat{\bf E}\parallel xy$, additional complex structures
are more apparently observed in experiment.
It is insisted that the system should consist of two components, 
both one-dimensional and two-dimensional characters even in
the static stripe phase in Nd-LSCO\cite{Zhou00}. 
$\widehat{\bf E}$-dependent behaviors in our calculations,
however, come from the interferences of noninteracting stripes (chains)
and may be expected not fully justified in the stripe system.
In the stripe system much two-dimensional characters induced by
stripe fluctuations and interactions of the stripes
with the two-dimensional background 
exist unlike a pure chain system.
This is the reason that our study is successful for PBCO,
whereas less successful in Nd-LSCO.

Finally, we may need to clarify the scope or limit for the present
photoemission approach. We have thought of the noninteracting
(noninteracting among chains and also between chains 
and the rest of the system) 
one-dimensional chains (stripes) for both PBCO and Nd-LSCO. Therefore,
in such a sense, the matrix effects addressed in our study
are purely geometric. This must be
quite drastic approximations, but can be well consistent with 
our original purpose of studying the matrix effects. 
Further investigations for the dipole matrix effect
are suggested to take into account better considerations
for the stripes (allowing the transverse excitations)
and thus leading to the surface Green's function,
which would make it possible to describe
the dynamical stripe phase too. We should also think of how good 
the plane wave for the photoelectron could be in these 
anisotropic systems.

\section*{acknowledgement}
One of the authors (J.D.L.) acknowledges the fellowship 
from the Japan Society for the Promotion of Science.
This work was supported by a Grant-in-aid for Scientific Research in 
the Priority Area "Novel Quantum Phenomena in Transition Metal Oxides"
from the Ministry of Education, Science, Sports, and Culture and
by the Special Coordination Fund for the Promotion of Science and Technology
from the Science and Technology Agency.

\end{multicols}
\end{document}